\documentclass{jps-cp}
\usepackage{txfonts} 

\title{muEDM: Towards a search for the muon electric dipole moment at PSI using the frozen-spin technique}

\author{Mikio~\textsc{Sakurai}$^{1,*}$, Andreas~\textsc{Adelmann}$^{1,2}$, Malte~\textsc{Backhaus}$^{1}$, Niklaus~\textsc{Berger}$^{3}$, Manfred~\textsc{Daum}$^{2}$, Kim~Siang~\textsc{Khaw}$^{4,5}$, Klaus~\textsc{Kirch}$^{1,2}$, Andreas~\textsc{Knecht}$^{2}$, Angela~\textsc{Papa}$^{2,6}$, Claude~\textsc{Petitjean}$^{2}$ and Philipp~\textsc{Schmidt-Wellenburg}$^{2,\dagger}$}

\inst{$^{1}$Institute for Particle Physics and Astrophysics, ETH Zürich, 8093 Zürich, Switzerland\\
	$^{2}$Paul Scherrer Institute, 5232 Viligen PSI, Switzerland\\
    $^{3}$PRISMA$^{+}$ Cluster of Excellence and Institute of Nuclear Physics, Johannes Gutenberg University Mainz, Mainz, Germany\\
    $^{4}$Tsung-Dao Lee Institute, Shanghai Jiao Tong University, Shanghai, China\\
    $^{5}$School of Physics and Astronomy, Shanghai Jiao Tong University, Shanghai, China\\
    $^{6}$Dipartimento di Fisica, University of Pisa and INFN Sezione di Pisa, Largo B. Pontecorvo 3, 56127 Pisa, Italy}

\email{$^*$msakurai@phys.ethz.ch, $^\dagger$philipp.schmidt-wellenburg@psi.ch}

\recdate{January 15, 2022}

\abst{The search for a permanent electric dipole moment (EDM) of the muon is an excellent probe for physics beyond the Standard Model of particle physics. We propose the first dedicated muon EDM search employing the frozen-spin technique at the Paul Scherrer Institute (PSI), Switzerland, with a sensitivity of $6 \times 10^{-23}$~$e\!\cdot\!\mathrm{cm}$, improving the current best limit set by the E821 experiment at Brookhaven National Laboratory by more than three orders of magnitude. In preparation for a high precision experiment to measure the muon EDM, several R\&D studies have been performed at PSI: the characterisation of a possible beamline to host the experiment for the muon beam injection study and the measurement of the multiple Coulomb scattering of positrons in potential detector materials at low momenta for the positron tracking scheme development. This paper discusses experimental concepts and the current status of the muEDM experiment at PSI.}

\kword{muon, EDM, frozen-spin technique, PSI}

\begin{document}
\maketitle

\section{Introduction}
We propose a search for the permanent electric dipole moment (EDM) of the muon employing the frozen-spin technique~\cite{Farley2004} at the Paul Scherrer Institute (PSI), Switzerland, aiming for a sensitivity of $6 \times 10^{-23}$~$e\!\cdot\!\mathrm{cm}$~\cite{Adelmann2021}. A non-zero EDM of a fundamental particle violates both parity and time-reversal symmetry, and assuming CPT invariance, this implies the violation of charge-parity (CP) symmetry. The Standard Model of particle physics (SM) predicts tiny EDMs, many orders of magnitude below current experimental limits, hence, any detection of non-zero EDMs strongly indicates new CP violation sources from physics beyond the SM (BSM). Since CP violation is one of the three Sakharov conditions to generate the matter-dominated Universe~\cite{Sakharov1967} and CP violation in the SM is insufficient to explain this, EDMs of fundamental particles are excellent probes of BSM to search for additional CP violation sources.

The muon is the only fundamental particle for which the EDM is directly probed on the bare particle. The muon EDM has been measured in conjunction with the muon $g-2$ in recent decades. The current experimental limit of $d_{\mu}<1.8 \times 10^{-19}$~$e\!\cdot\!\mathrm{cm}$ (95\% C.L.) is also based on an analysis of the $g-2$ data of the BNL E821 experiment~\cite{Bennett2009}. By considering linear mass scaling of the electron and muon, the current electron EDM limit of $d_{e}<1.1 \times 10^{-29}$~$e\!\cdot\!\mathrm{cm}$ (90\% C.L.)~\cite{Andreev2018} implies an indirect muon EDM limit of about 8 orders of magnitude smaller than the current limit. However, the recently reported combined $4.2\sigma$ discrepancy in the muon $g-2$~\cite{Abi2021} as well as the observed tensions in $B$-decays~\cite{Aaij2013a,Aaij2014,Aaij2015,Aaij2016,Aaij2017,Aaij2021} motivate BSM models which could allow a larger muon EDM without tight constraints from other lepton EDMs~\cite{Crivellin2018}. Therefore, the dedicated muon EDM search plays an important role on further pushing EDM searches beyond the first generation fundamental particles and probing the role of the lepton flavour universality in nature.

\section{The muon EDM experiment at PSI}
Similarly to the magnetic dipole moment, the EDM can be defined as  $\vec{d_{\mu}}=\eta \frac{e}{2m_{\mu}c}\vec{s}$, where $\eta$ is the size of the EDM and $e$, $m_{\mu}$ and $\vec{s}$ are the charge, mass and spin vector of the muon, respectively. The muon spin precession $\vec{\omega}$ in the presence of a magnetic field $\vec{B}$ and an electric field $\vec{E}$ can be expressed as
\begin{equation}
	\vec{\omega}=	\vec{\omega_{a}}+	\vec{\omega_{e}}=-\frac{e}{m_{\mu}}\left[ \left\lbrace  a_{\mu}\vec{B}-\left( a_{\mu}+\frac{1}{1-\gamma^{2}}\right) \frac{\vec{\beta}\times\vec{E}}{c}\right\rbrace  +\frac{\eta}{2}\left\lbrace \vec{\beta}\times\vec{B}+\frac{\vec{E}}{c}\right\rbrace \right] ,
	\label{TBMT}
\end{equation}
where $\vec{\omega_{a}}$, corresponding to the first brace term, and $\vec{\omega_{e}}$, corresponding to the second brace term, are spin precessions due to $g-2$ and EDM, respectively. Here, $a_{\mu}=(g-2)/2$ is the anomalous magnetic moment of the muon and $\gamma=1/\sqrt{1-\beta^{2}}$ is the Lorentz factor. The BNL E821 experiment chose the ``magic'' momentum of $p_{magic}=m_{\mu}/\sqrt{a_{\mu}}=3.09$~GeV/$c$, such that $a_{\mu}+1/(1-\gamma^{2})=0$, to cancel the effect of the focusing electric field on the muon spin precession. Hence, if the muon has a non-zero EDM, the EDM term causes a tilt of the spin precession plane which leads to a vertical oscillation $\pi/2$ out of phase with respect to the horizontal precession due to the $g-2$, increasing the observed precession frequency to $\omega=\sqrt{\omega_{a}^{2}+\omega_{e}^{2}}$.

\begin{figure}[tbh]
	\begin{center}
		\begin{tabular}{cc}
			\begin{minipage}{0.5\columnwidth}
				\begin{center}
					\includegraphics[width=0.7\textwidth]{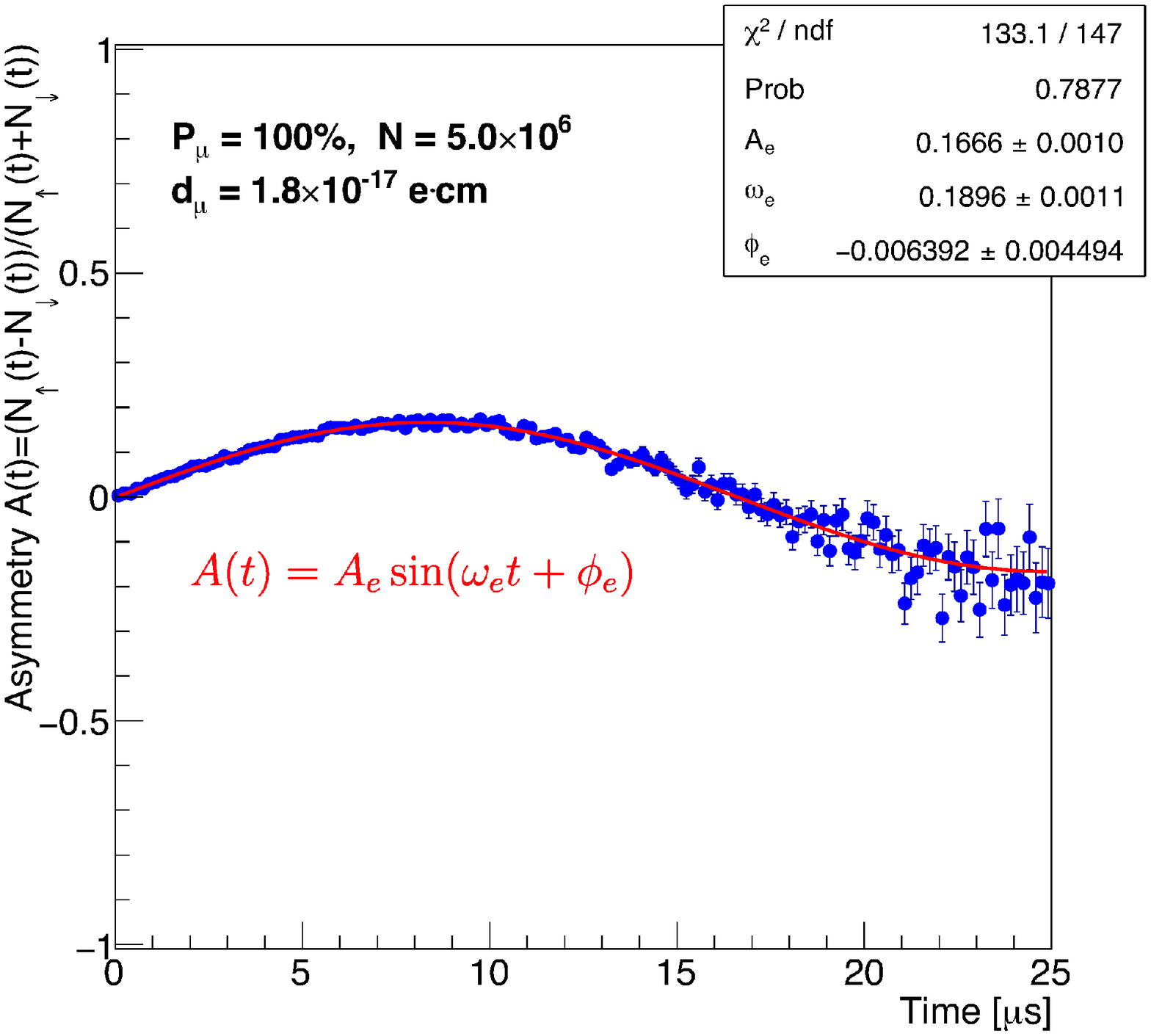}\\
				\end{center}
			\end{minipage}
			\begin{minipage}{0.5\columnwidth}
				\begin{center}
					\includegraphics[width=0.7\textwidth]{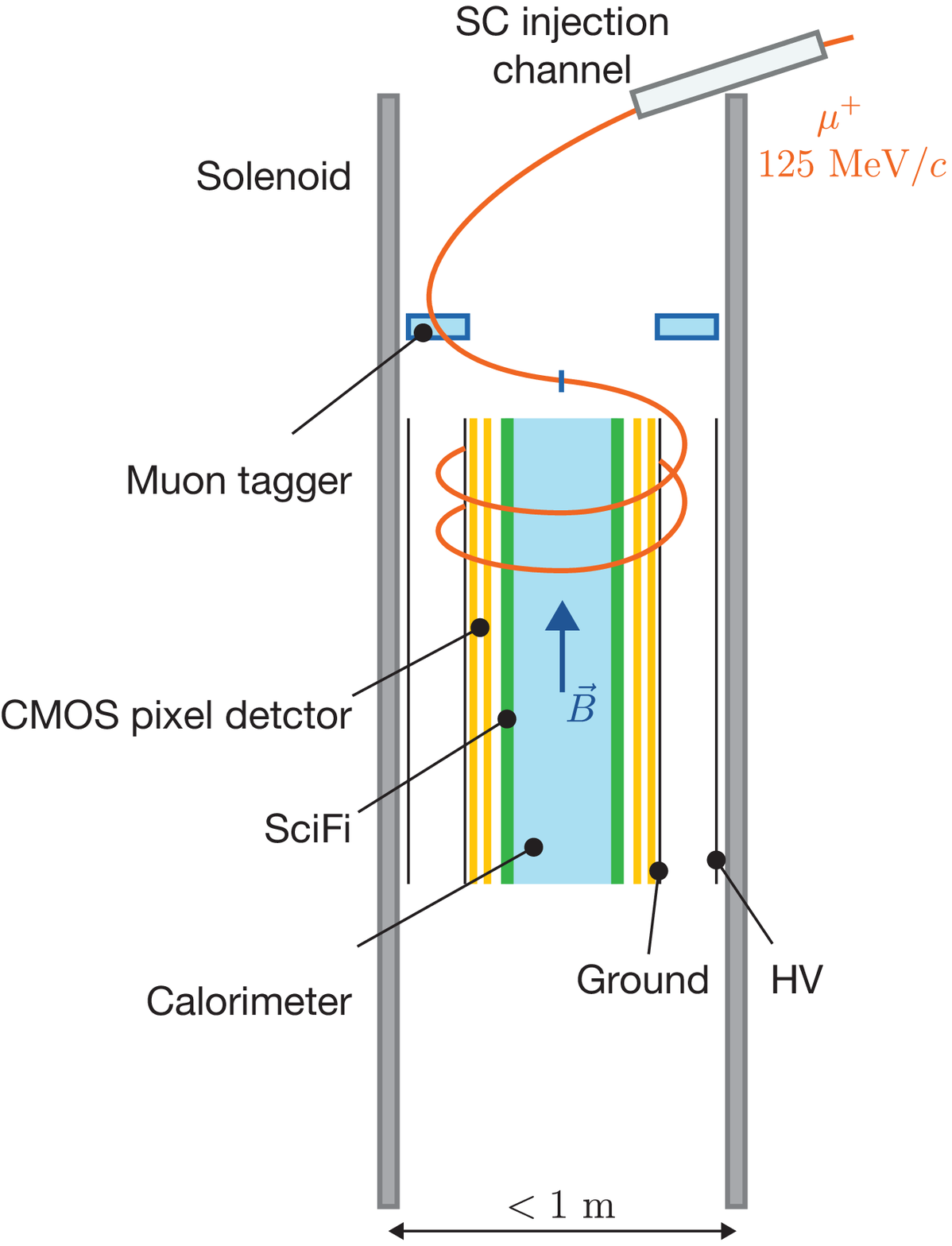}\\
				\end{center}
			\end{minipage}
		\end{tabular}
		\caption{(Left) Simulated ideal positron up-down counting asymmetry as a function of time assuming a large muon EDM of $d_{\mu}=1.8 \times 10^{-17}$~$e\!\cdot\!\mathrm{cm}$. (Right) Concept of the muon EDM experiment at PSI.}
		\label{PSImuEDM}
	\end{center}
\end{figure}

The muon EDM experiment at PSI is the first dedicated muEDM search based on the concepts discussed in~\cite{Farley2004,Adelmann2010,Iinuma2016}. The key concept of the experiment is the usage of the frozen-spin technique to purely observe the spin precession $\vec{\omega_{e}}$ caused by the EDM coupling to the motional electric field, $\vec{E^{*}}=\gamma c \vec{\beta}\times\vec{B}$, in the muon rest frame while cancelling the $g-2$ precession in Eq.~(\ref{TBMT}) by applying a radial electric field $E_\mathrm{f}\approx a_{\mu}Bc\beta\gamma^{2}$. Thus, the muon spin is ``frozen'' to the muon momentum vector if $\eta=0$. In the presence of the EDM, Eq.~(\ref{TBMT}) simplifies to
\begin{equation}
	\vec{\omega}=\vec{\omega_{e}}=-\frac{e\eta}{2m_{\mu}}\left[ \vec{\beta}\times\vec{B}+\frac{\vec{E_\mathrm{f}}}{c}\right] ,
	\label{TBMT-FS}
\end{equation}
resulting in a vertical build-up of the polarisation with time $t$. Therefore, the EDM signal can be measured as the up-down counting asymmetry of the muon decay positrons $A(t)=\alpha P_{0}\sin(\omega_{e}t)\approx\alpha P_{0}\omega_{e}t$, where $\alpha$  is the mean analysis power of the final polarisation and $P_{0}$ is the initial muon polarisation. Figure~\ref{PSImuEDM} left shows a simulated ideal up-down asymmetry of decay positrons as a function of time assuming an unphysically large muon EDM of $d_{\mu}=1.8 \times 10^{-17}$~$e\!\cdot\!\mathrm{cm}$.
The sensitivity of the muon EDM is then given by
\begin{equation}
	\sigma(d_{\mu}) =\frac{a_{\mu}\hbar\gamma}{2P_{0}E_\mathrm{f}\sqrt{N}\tau_{\mu}\alpha}, 
	\label{sensitivity}
\end{equation}
where $N$ is the number of detected decay positrons and $\tau_{\mu}$ is the muon lifetime.

Figure~\ref{PSImuEDM} right shows a schematic of the experimental concept for the muon EDM search at PSI. Polarised positive muons at $p=125$~MeV/$c$ ($\gamma=1.77$) with $P_{0}\approx95\%$ from the PSI $\mu$E1 beamline are injected vertically from the top into a compact storage solenoid with $B=3$~T through a superconducting injection channel. A magnetic kicker at the centre of the solenoid is triggered by the fast scintillator signal in the injection channel to guide muons onto a stable orbit. The radial electric field $E_\mathrm{f}=2$~MV/m to cancel the $g-2$ spin precession is generated by two concentric cylindrical electrodes. The muons are stored one at a time at a rate of about $60$~kHz and the decay positrons are detected by positron tracking systems made of CMOS pixel detectors and scintillating fibres for precise timing. The muon tagger is for the decay vertex reconstruction and provides a measurement of time and trajectory of the muon at the point of injection into the magnetic field. An option to use a calorimeter is also considered for measuring positron energy. Assuming $N\sim10^{12}$ per year and from Eq.~(\ref{sensitivity}) together with the mean analysis power of $\alpha=0.3$, a statistical sensitivity of $\sigma( d_{\mu}) \approx 6 \times 10^{-23}$~$e\!\cdot\!\mathrm{cm}$ can be achieved in one year of data taking at the PSI $\mu$E1 beamline.

\section{R\&D studies at PSI}
\subsection{Characterisation of the potential beamline}
The precision muon EDM measurement requires a high-flux polarised muon beam with a fairly high momentum as its sensitivity scales with $\sqrt{N}$, $P_{0}$ and $\gamma$ from Eq.~(\ref{sensitivity}) with $E_\mathrm{f}\approx a_{\mu}Bc\beta\gamma^{2}$. This favours the $\mu$E1 beamline at PSI  as a potential beamline to host the experiment in order to achieve the highest sensitivity to a muon EDM measurement. Therefore, in 2019, a characterisation of the $\mu$E1 beamline was performed to obtain essential input parameters for simulations towards the experiment as well as the injection study. The muon beam rate, transverse phase space (emittance) and polarisation level were studied for several muon momenta from $65$~MeV/$c$ up to $125 $~MeV/$c$. 

\begin{figure}[h]
	\begin{center}
		\begin{tabular}{cc}
			\begin{minipage}{0.33\columnwidth}
				\begin{center}
					\includegraphics[width=\textwidth]{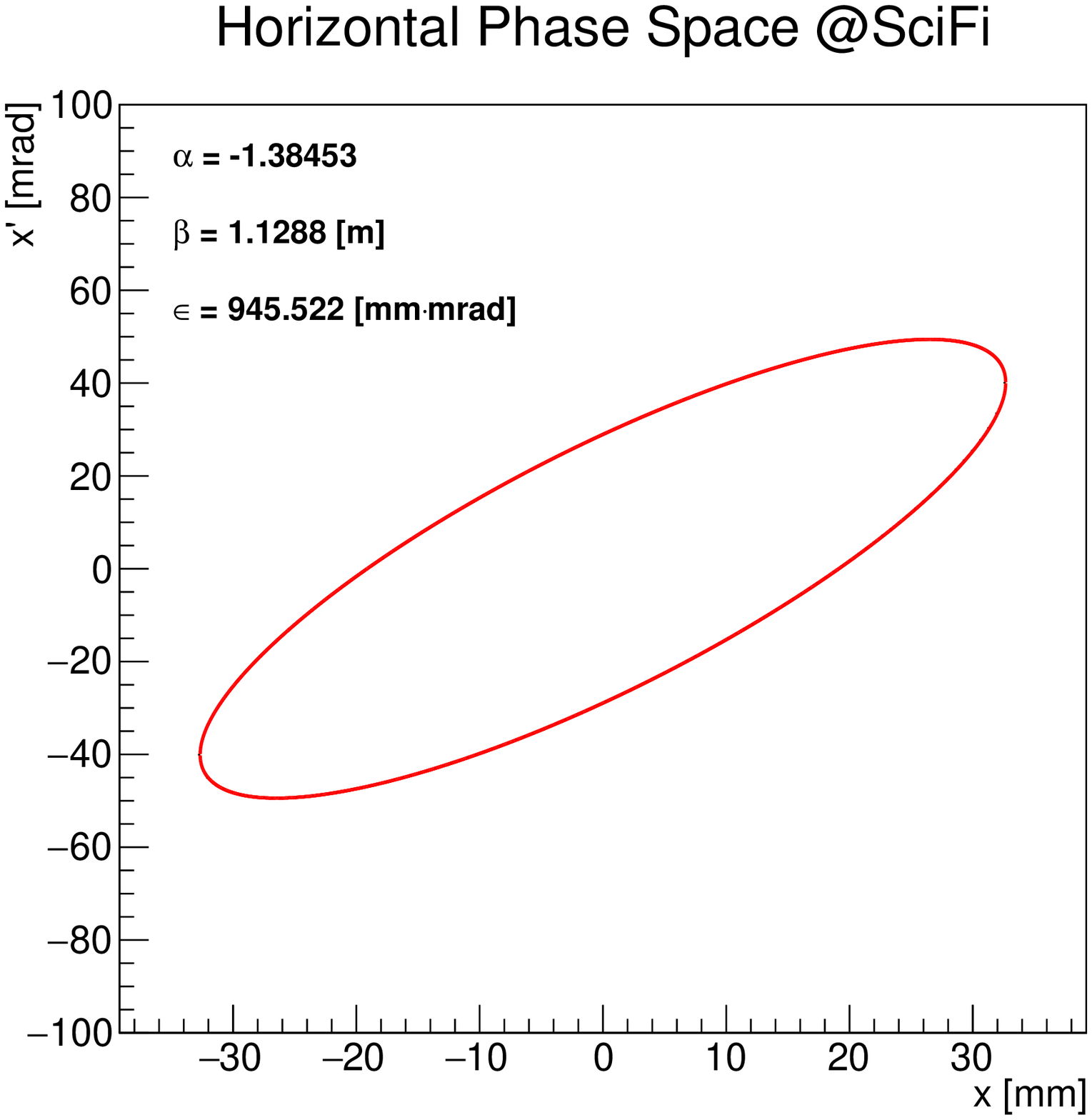}\\
				\end{center}
			\end{minipage}
			\begin{minipage}{0.33\columnwidth}
				\begin{center}
					\includegraphics[width=\textwidth]{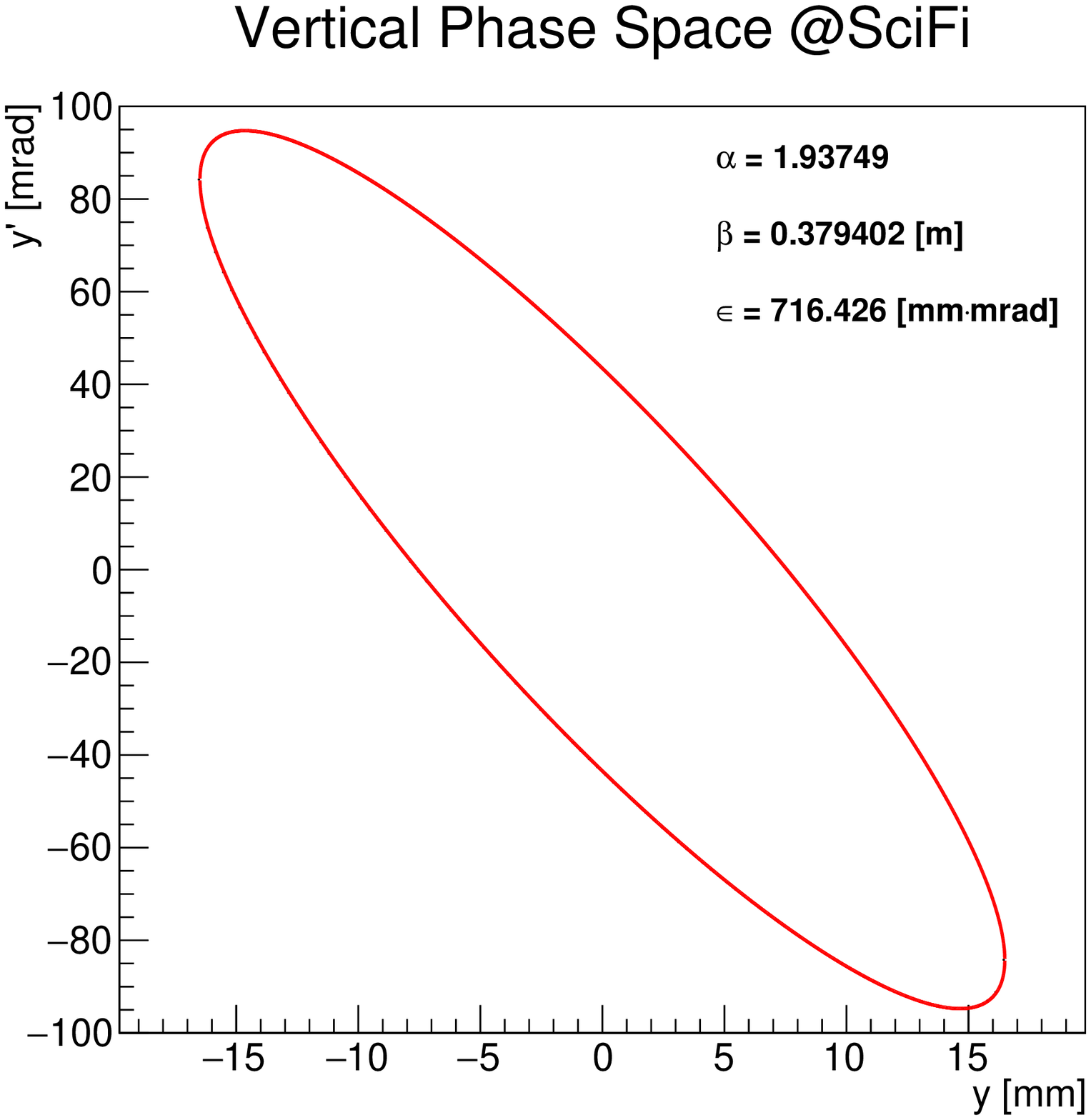}\\
				\end{center}
			\end{minipage}
			\begin{minipage}{0.33\columnwidth}
				\begin{center}
					\includegraphics[width=\textwidth]{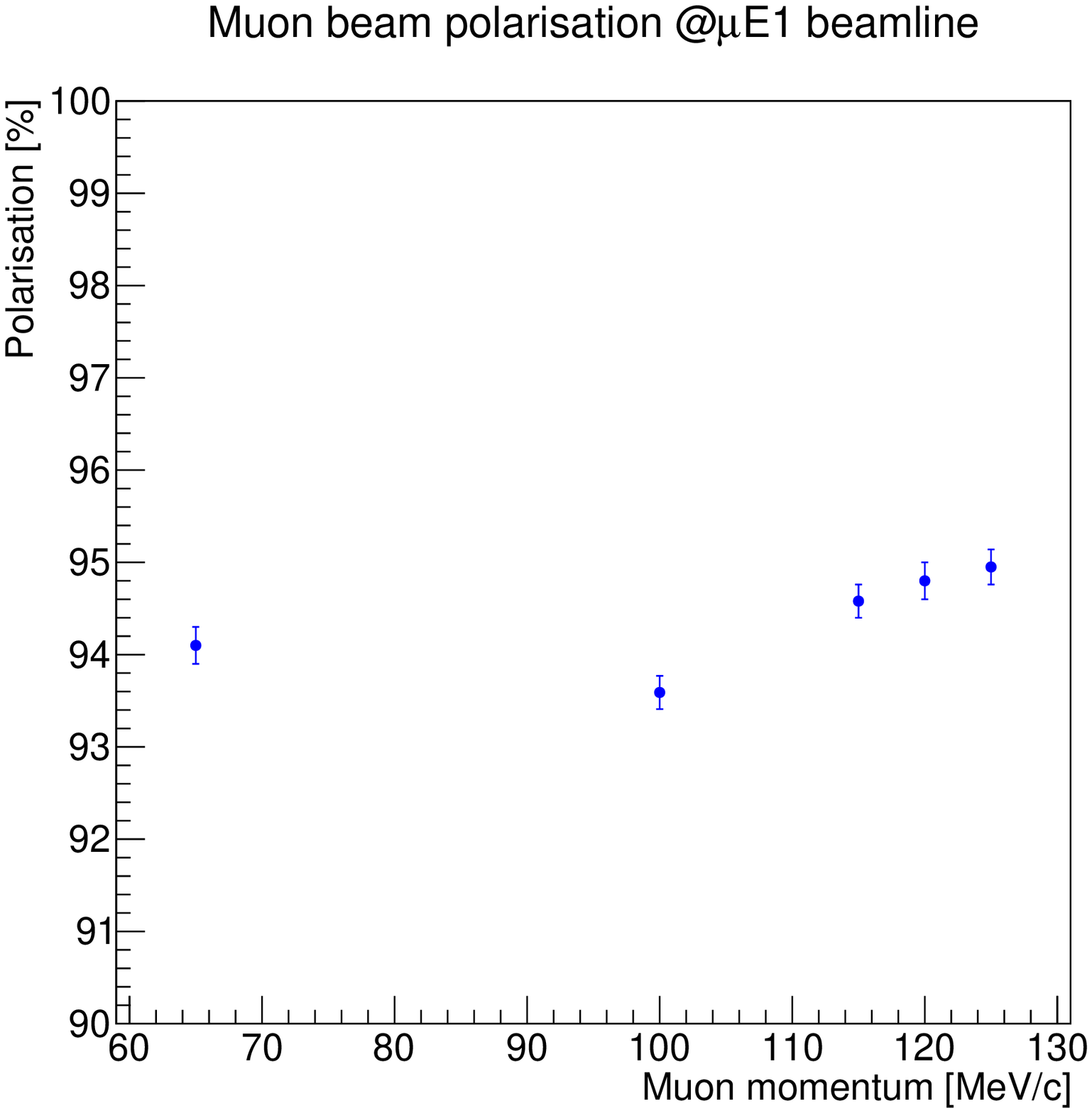}\\
				\end{center}
			\end{minipage}
		\end{tabular}
		\caption{Horizontal (left) and vertical (middle) $1\sigma$ phase space ellipses of  $125$~MeV/$c$ muon beam  at the SciFi beam monitor position at the PSI $\mu$E1 beamline. (Right) Muon beam polarisation level at the PSI $\mu$E1 beamline as a function of the muon beam momentum.}
		\label{muE1_Characterisation}
	\end{center}
\end{figure}

A scintillating fibre (SciFi) beam monitor~\cite{Papa2020} mounted $526$~mm downstream of the final quadrupole of the beamline was used to measure the beam rate and transverse beam size. The transverse phase space was extracted using a quadrupole scan technique. A maximum muon beam rate of $8.0\times10^{7}$~$\mu^{+}/$s at $1.6$~mA proton current, which scales to $1.2\times10^{8}$~$\mu^{+}/$s for the nominal proton accelerator operation current of $2.4$~mA, was achieved at $125$~MeV/$c$ and the corresponding horizontal and vertical phase space ellipses with emittances of $945$~mm$\cdot$mrad and $716$~mm$\cdot$mrad ($1\sigma$), respectively, are shown in Fig.~\ref{muE1_Characterisation} left and middle. The polarisation of the beam was also measured using a copper target to stop muons inside the existing $\mu$SR detector of the GPD instrument~\cite{Khasanov2016} at the $\mu$E1 beamline. An absolute muon beam polarisation was extracted by comparing the measurement with Geant4 simulations assuming $100\%$ beam polarisation. Figure~\ref{muE1_Characterisation} right shows the muon beam polarisation level as a function of the muon beam momentum. The polarisation of the muon beam is $(94.95\pm0.19)\%$ at $125$~MeV/$c$.

\subsection{Measurement of the multiple Coulomb scattering of positrons at low momenta}
\begin{figure}[h]
	\begin{center}
		\begin{tabular}{cc}
			\begin{minipage}{0.5\columnwidth}
				\begin{center}
					\includegraphics[width=0.67\textwidth]{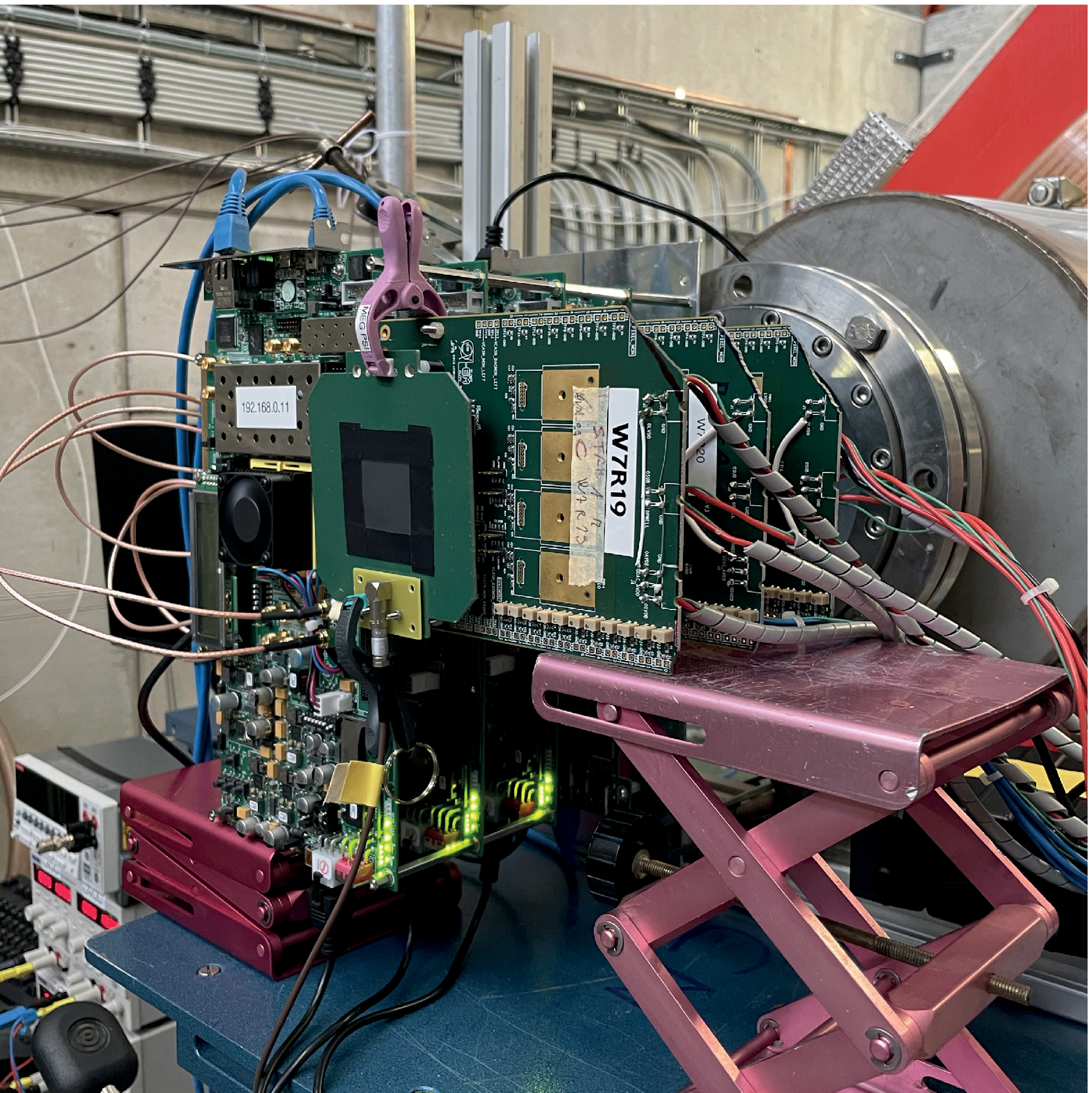}\\
				\end{center}
			\end{minipage}
			\begin{minipage}{0.5\columnwidth}
				\begin{center}
					\includegraphics[width=\textwidth]{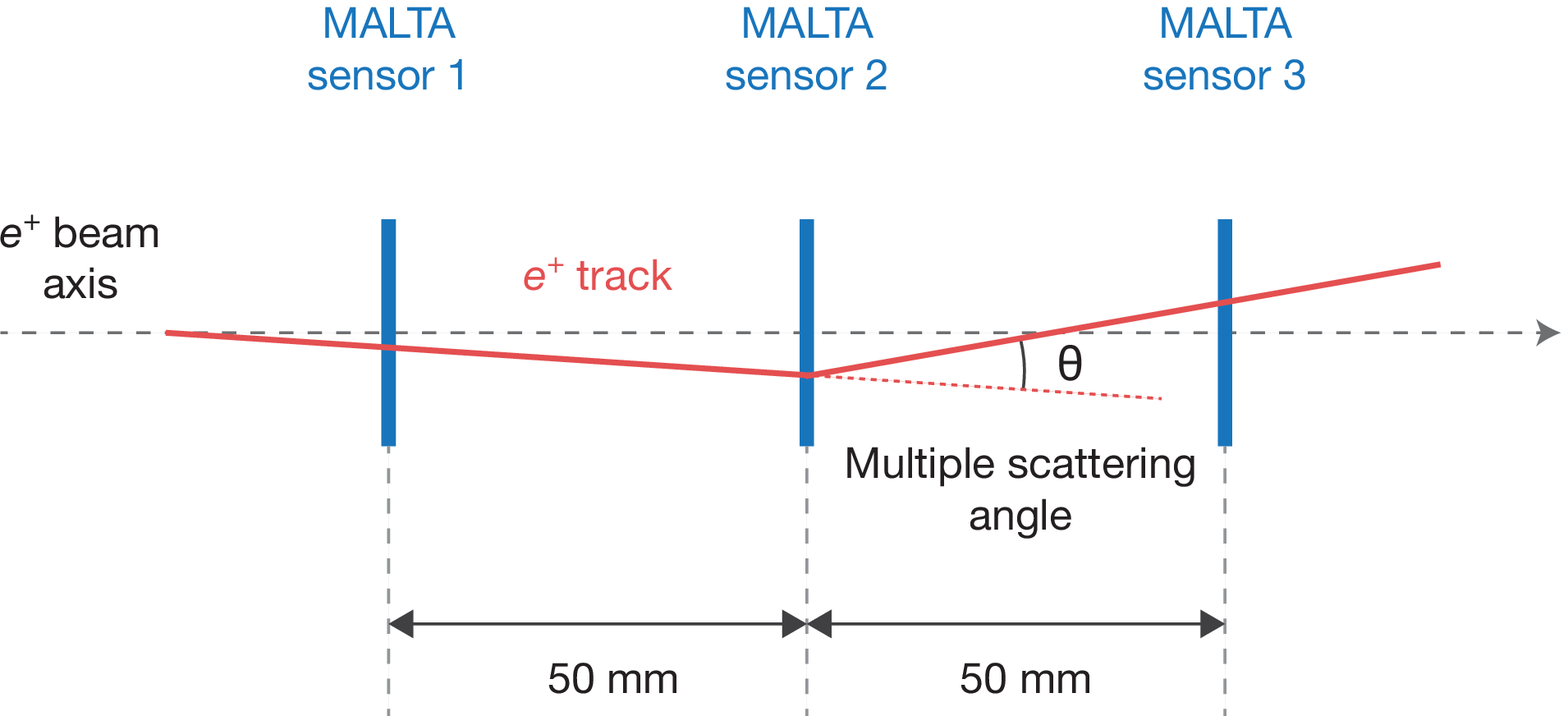}\\
				\end{center}
			\end{minipage}
		\end{tabular}
		\caption{Picture (left) and schematic (right) of the 3-plane MALTA telescope used for the multiple scattering measurement. In the left picture, the positron beam enters from the right.}
		\label{MALTAsetup}
	\end{center}
\end{figure}

Multiple Coulomb scattering is a series of small-angle scatterings which particles experience when passing through a material. This effect becomes dominant at low beam momenta, hence, the particle tracking resolution would be limited by the multiple scattering in materials. This strongly influences the positron tracking scheme as well as the detector design of the muEDM experiment. Therefore, a very good understanding of the multiple scattering in potential detector materials is crucial to establish experimental concepts. In 2020, the multiple Coulomb scattering of positrons in possible detector materials, namely a single MALTA sensor (a silicon of $307.8$~$\mu$m thickness)~\cite{Cardella2019}, and $500$~$\mu$m thick Kapton and Mylar foils, was measured at positron beam momenta from $20$~MeV/$c$ to $85$~MeV/$c$ at the PSI $\pi$E1 beamline using a 3-plane telescope based on the MALTA pixel detector. Figure~\ref{MALTAsetup} left and right present a picture and schematic of the 3-plane MALTA telescope used for the multiple scattering measurement, respectively. 

The multiple scattering angle is often parametrised with the so-called Highland formula~\cite{Highland1975a,Lynch1991} and evaluated as the RMS of the central $98\%$ of the projected angular distribution. For the multiple scattering measurement in the MALTA sensor, the second plane of the telescope was used as an active scattering target. By selecting tracks which have exactly 1 pixel hit per telescope plane, the angle between tracks from the first to second and the second to third telescope plane was calculated in both horizontal and vertical projections as a multiple scattering angle $\theta$. Similar measurements were performed by placing either a $500$~$\mu$m thick Kapton or Mylar foil $10$~mm downstream of the second plane as a scattering target. The multiple scattering in those foils was deconvolved from the measured angle $\theta$ with the foil by assuming it is a convolution of the scattering contributions from both telescope plane and the foil. A reasonable description of the measured multiple scattering angle is confirmed by the Highland formula as well as several multiple scattering models available in the Geant4 simulation. 

\section{Conclusions}
The search for the muon EDM is a unique opportunity to explore CP violation in BSM physics as well as to probe lepton flavour universality in one of the least tested areas of the SM. The first dedicated muon EDM experiment employing the frozen-spin technique at PSI aims to achieve a sensitivity of $6 \times 10^{-23}$~$e\!\cdot\!\mathrm{cm}$, more than three orders of magnitude better than the current best limit set by the BNL E821 experiment. Towards a high precision muon EDM measurement, the potential beamline characterisation and a study of the multiple Coulomb scattering of positrons at low momenta were performed at PSI. The letter of intent for the muEDM experiment was submitted to the PSI CHRISP research committee in 2021 together with collaborators from 20 different institutions~\cite{Adelmann2021} and several R\&D studies are currently underway for the full experimental proposal submission to PSI.

\section*{Acknowledgements}
We would like to thank H.~Pernegger and C.~Solans~Sanchez for their support and providing the MALTA sensors for the multiple scattering measurement. This work was supported by ETH Research Grant ETH-48 18-1.

\end{document}